\documentclass[twocolumn,showpacs,preprintnumbers,amsmath,amssymb,nofootinbib]{revtex4}

\usepackage{graphicx}
\usepackage{dcolumn}
\usepackage{bm}

\begin{document}

\title{Detection of the  Natural Alpha Decay of Tungsten\\}

\author{C. Cozzini$^{1,}$\footnote{Corresponding author. \emph{Email
address:} c.cozzini@physics.ox.ac.uk}, G. Angloher$^2$, C. Bucci$^3$,
F. von Feilitzsch$^4$, D. Hauff$^2$, S. Henry$^1$,\\ Th. Jagemann$^4$,
J. Jochum$^5$, H. Kraus$^1$, B. Majorovits$^1$, V. Mikhailik$^1$,
J. Ninkovic$^2$,\\ F. Petricca$^2$, W. Potzel$^4$, F. Pr\"{o}bst$^2$,
Y. Ramachers$^{1,}$\footnote{Present address: University of Warwick,
Coventry CV4 7AL, U.K.}, W. Rau$^4$, M. Razeti$^4$,\\ W. Seidel$^2$,
M. Stark$^4$, L. Stodolsky$^2$, A.J.B. Tolhurst$^1$, W. Westphal$^4$,
H. Wulandari$^4$}

\affiliation{$^1$Department of Physics, University of Oxford, Oxford OX1
3RH, U.K.\\$^2$MPI f\"{u}r Physik, F\"{o}hringer Ring 6, 80805 Munich,
Germany\\$^3$Laboratori Nazionali del Gran Sasso, 67010 Assergi,
Italy\\$^4$Physikdepartment E-15, TU M{\"u}nchen, James-Franck-Str.,
85748 Garching, Germany\\$^5$Eberhard-Karls-Unversit{\"a}t T{\"u}bingen, D-72076
T{\"u}bingen, Germany}

\date{\today}

\begin{abstract}
 The natural $\alpha$-decay of $^{180}$W has been
unambiguously detected  for the first time. The $\alpha$ peak is found in a ($\gamma ,\beta$ and neutron)-free background
spectrum. This has been achieved by the simultaneous measurement of
phonon and light signals with the CRESST cryogenic detectors. A
half-life of T$_{1/2} = (1.8 \pm 0.2) \times 10^{18}$ y and an energy
release of
Q = (2516.4 $\pm$ 1.1 (stat.) $\pm$ 1.2 (sys.)) keV have been
measured. New limits are also set on the half-lives of the other
naturally occurring tungsten isotopes.
\end{abstract}

\maketitle

\section{Introduction}
 The  $\alpha$-decay of the naturally occurring isotopes of tungsten (W)
has been the subject of experimental search for many decades.
Tungsten is an interesting element because $\alpha$-decay is
energetically allowed for all five naturally occurring
isotopes. 
Mass
excess measurements \cite{wapstra} show that the available decay energy  Q for
all these isotopes is  low ($<3$ MeV) and that the Q values lie in the same energy
range as  $\beta$ and $\gamma$ decay due to natural chains (see Table
\ref{Qvalues}). Therefore background suppression and possibly event-by-event
discrimination against $\gamma$'s and $\beta$'s, which may obscure the $\alpha$ signal,  are
crucial issues for the detection of such rare events.
\begin{table}
\begin{center}
\begin{tabular}{|l|rr|rr|}
\hline
Isotope&\multicolumn{2}{c|}{Abundance[$\%$]}&\multicolumn{2}{c|}{Q [keV]}\\
&IUPAC \cite{IUPAC}& Ref.\footnote{Considered by
the IUPAC (International Union of Pure and Applied Chemistry) to be
the best single measurements.}
\cite{Volkening}&Ref. \cite{wapstra}&Ref. \cite{wapstra03} \\\hline
$^{180}$W&0.12(1)&0.1198(2)&2516(5)&2508(4)\\
$^{182}$W&26.50(16)&26.4985(49)&1774(3)&1771.8(2.2)\\
$^{183}$W&14.31(4)&14.3136(6)&1682(3)&1680.0(2.2)\\
$^{184}$W&30.64(2)&30.6422(13)&1659(3)&1656.2(2.2)\\
$^{186}$W&28.43(19)&28.4259(62)&1123(7)&1124(7)\\
\hline
\end{tabular}
\caption{Representative Isotopic Composition according to the IUPAC, best single
measurements on isotopic abundance from  individual references, and decay energies for the isotopes
of natural tungsten.}
\label{Qvalues}
\end{center}
\end{table}

While
$^{182}$W,$^{183}$W,$^{184}$W, and $^{186}$W are expected to have
half-lives well above $10^{32}$ y, the isotope $^{180}$W is expected to have  a half-life near $10^{18}$ y, and is close to observability with
present techniques.
Most recently, the Kiev-Firenze
Collaboration, operating  CdWO$_4$ scintillation detectors    
in the Solotvina mine
reported a ``first indication of a possible $\alpha$-decay of
$^{180}$W'' with a half-life of
T$_{1/2} = 1.1^{+0.8}_{-0.4} (stat.) \pm 0.3 (sys.) \times 10^{18}$ years \cite{kiev}.  
 It is well
known \cite{birks} that the scintillation yield for $\alpha$ particles is
lower than that for $\beta$ or $\gamma$ particles of the same
energy. After pulse shape analysis, an $\alpha$-peak at $\sim300$ keV  was interpreted as 
$^{180}$W $\alpha$-decay. Due to the low energy resolution (FWHM = 110 keV) it
was impossible to exclude alternative
explanations for the peak, hence the result was
treated as an indication. They also
set   a 90$\%$ C.L. lower limit of T$_{1/2} \geq 0.7 \times 10^{18}$
years.  Also the ROSEBUD collaboration,
operating a 54 g CaWO$_4$ scintillating bolometer at the Canfranc
Underground Laboratory, obtained a  limit of T$_{1/2} \geq 1.7
\times 10^{17}$ years at 90$\%$ C.L \cite{cebrian}.

\begin{figure*}
\begin{center}
\includegraphics[height=9cm]{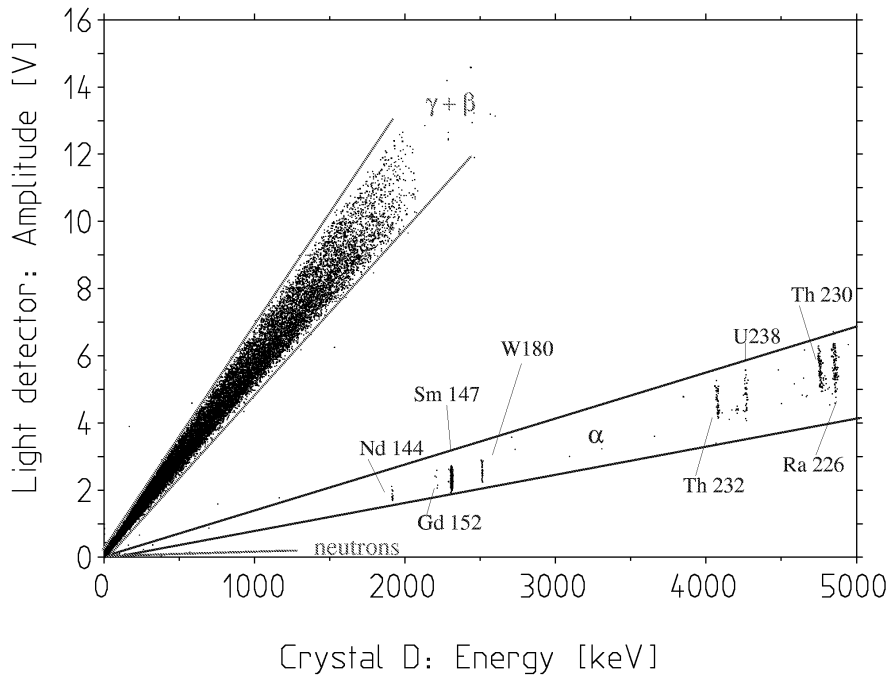}
\caption{\small Pulse amplitude in the light channel versus energy in
the  CaWO$_4$ crystal D from run 28. Three clearly separated event
populations appear. The solid lines border the three different
bands: the upper band is  due to $\gamma$'s and $\beta$'s, the lowest one
to neutrons and the middle one is due to $\alpha$'s. Each peak in the $\alpha$
band has been identified (see text) and is here labelled accordingly.}
\label{d2u_g}
\end{center}
\end{figure*} 
The high sensitivity and excellent energy  resolution of  low temperature
detectors (see for example \cite{cresstI,fiorini} ) and the great efforts
made to suppress and understand radioactive background
 in  dark matter searches and double beta decay experiments
 make such apparatus  sensitive to rare
nuclear decays. For example, the $\alpha$-decay of $^{209}$Bi
(considered as the heaviest stable isotope) has been recently detected
by a French group developing low temperatures bolometers for Dark
Matter direct detection \cite{marcillac}.
CRESST (Cryogenic Rare Event Search with Superconducting Thermometers) is a low background cryogenic facility
primarily devoted to the direct detection of  WIMP dark matter particles via their scattering by nuclei. Such nuclear recoils resulting from WIMP interaction can
be discriminated from electron background (caused by photons or
electrons) by measuring phonons and scintillation light
simultaneously. 
The energy detected via phonons in a cryogenic detector is, to
first order, independent of the nature of the particle. There is, however, a
significant difference in scintillation yield for nuclei and
electrons of the same energy. The
fraction of energy released in the nuclear interaction channel, negligible for
photons and electrons, becomes important for $\alpha$'s and heavy
ions and is dominant for WIMPs and neutrons.
As a result WIMP and neutron induced recoils give
considerably less scintillation light than electrons of the
same energy while $\alpha$ particles
can be clearly discriminated 
since they interact partly with both, nuclei and electrons.
This results in a pure $\alpha$ spectrum, i.e. one without contributions from
$\beta, \gamma$ and neutron events as in the middle band of Fig. \ref{d2u_g}.
Finally we remark that the energy measured following an internal $\alpha$-decay in a cryogenic
detector corresponds to the sum of the energies of the $\alpha$ particle and
of the recoiling nucleus, i.e. the total decay energy Q.  

In this paper we present clear evidence for  the $\alpha$-decay of
$^{180}$W from early runs of the CRESST II dark matter detectors, originating
from the
tungsten in the  CaWO$_4$ crystals  used as the dark matter target.

\section{Experimental set-up}
\begin{figure}
\begin{center}
\includegraphics[height=5cm]{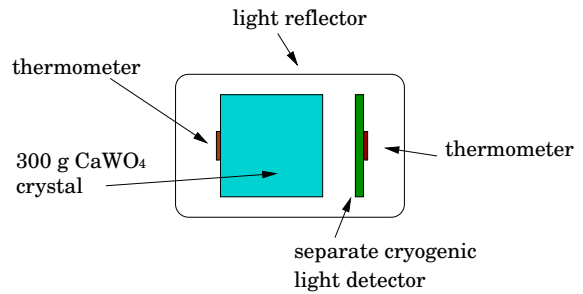}
\caption{\small Schematic of a detector module. It consists of a
scintillating 300 g CaWO$_4$ crystal (phonon channel) and a Si wafer
(light channel), both read out by a SPT. The set-up is surrounded by
a reflective foil}
\label{module}
\end{center}
\end{figure}
CRESST is an ultra-low background facility operating at the Gran Sasso
underground laboratories. Detailed descriptions can be found in
\cite{cresstI,cresstII}.
CRESST  has developed very sensitive cryogenic detectors consisting of
a dielectric target crystal with a superconducting phase transition	
thermometer (SPT)
evaporated onto one surface. 
Particle interactions in a cryogenic detector produce
phonons which propagate to the surface, where they heat the
electron system in the SPTs. These
thermometers, biased in the middle of their superconducting-to-normal
transition near a temperature of  10 milli-kelvin, transform the
temperature rise of their electrons into a relatively large
increase of the resistance of the film. The resistance change of the
low impedance thermometer is then measured via a SQUID. The resistance increase is a measure for the energy deposited. 


The set-up for the simultaneous detection of scintillation light and
phonons is described in \cite{cresstII} and shown schematically in
Fig. \ref{module} for one absorber module.
It consists of two
independent detectors, each one with its own SPT with SQUID readout.
The main detector consists of a $\sim 300$g cylindrical CaWO$_4$ crystal
($\varnothing = 40$ mm, $h = 40$ mm). 
CaWO$_4$ is a well known scintillator with high light yield 
and very broad emission spectrum peaked at 420 nm. It is characterized
by a high  index of refraction (n = 1.92) and a decay time of 17 $\mu$s at 77
 K.
The scintillation light produced in each target crystal is detected
via an associated calorimeter consisting of a silicon wafer of (30
$\times 30 \times 0.45) $mm$^3$ volume
with a
20 nm thick SiO$_2$ layer on both surfaces. In CaWO$_4$ only a few percent of the absorbed energy is transformed into light. To
minimise light losses the whole module is thus enclosed in a highly
reflective foil.

In the data to be presented here, three different CaWO$_4$ crystals
(referred to as crystal B, D and E) were used in the Gran
Sasso set-up for a total of four different CRESST runs (22, 23,
27, 28). For detector operation, the
temperature of the thermometer  is controlled by a dedicated heater.
Additionally the heater is used to inject test pulses which monitor
the long term stability of the detector \cite{cresstII}. In particular,
in the runs 27 and 28 a large voltage pulse was produced, reaching the saturation
region of the superconducting transition specifically to monitor the detector response above 1 MeV. This pulse
was sent, at a rate of 0.5 Hz, throughout the  measuring period.
A record length of 4096 samples and a time base of 40 $\mu$s,
resulting in a time window of 164 ms for recording an event, was
chosen for all the runs presented.

\section{Off line analysis}

\begin{figure*}
\begin{center}
\includegraphics[height=5cm]{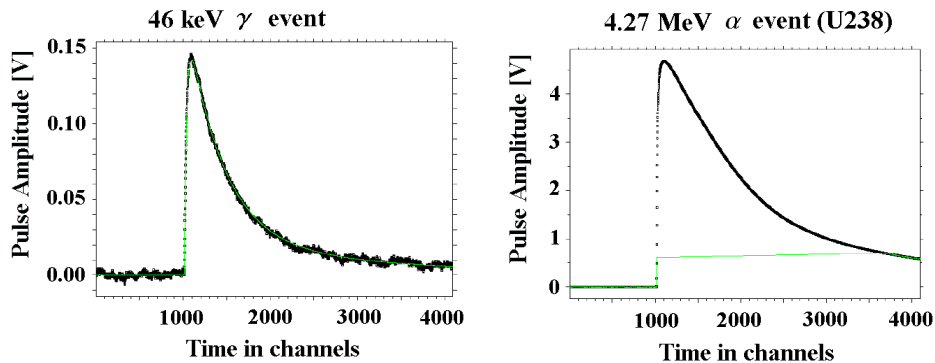}
\caption{\small Fit of two events from crystal D (run 28) with a
template obtained by averaging pulses from the 122 keV peak. The fit is truncated at 0.7
V. For pulses of amplitude below this value the fit procedure is
standard (left plot). For pulses with an amplitude exceeding this value,
the higher part of the pulse shape (above the horizontal dotted line
in the right plot) is distorted due to saturation.
Therefore the fit is truncated, i.e. fit only to data points below 0.7 V,  and the amplitude of the pulse is
reconstructed from the template.}
\label{truncatedfit}
\end{center}
\end{figure*}
 
CRESST  detectors are optimised  to be most sensitive in the energy
region relevant for dark matter direct detection ($<$200
keV). In this energy range the
detector  response is mapped out with electric heater test pulses, and the
122 keV line from a $^{57}$Co source provides the calibration of the test
pulses in terms of a $\gamma$ equivalent energy \cite{cresstI}.
 A pulse shape
template is obtained by averaging pulses from the 122 keV line. This template is used in
a fit to the actual signal pulses in order to
accurately determine the signal amplitude. This is valid as long as the detector response is in a
linear regime. If the response becomes non-linear the pulse shape
changes and the same template cannot be used over the full energy range. 

Typically, the pulse shape for the CaWO$_4$ detectors can be described by $\tau_{rise}$ = 1.1 ms and $\tau_{decay}$ = 30 ms.
However, the temperature rise
caused by $\alpha$-decays is usually beyond the dynamic range of the
SPT and for precise energy determination the signal
duration rather than the pulse height must be used. Therefore two new analysis
methods, described in \cite{methods}, 
were developed in order to reconstruct
the higher energies (MeV region), for which
the detector response is non-linear.

In the first method (Method A) a template from pulses in the
linear region (for example the 122 keV template from the calibration) is  used to fit  all
measured pulses with an amplitude below the voltage at which the response becomes non-linear. For all pulses with an amplitude exceeding that voltage the fit is truncated and the upper part is reconstructed from the template. This is shown in
Fig. \ref{truncatedfit}. The resulting amplitude spectrum is then linear by
construction.

If the pulse duration is too long the 
information from the decaying part of the pulse is outside the
recording time window and the method fails.  To overcome this
limitation a second method was developed (Method B). The template
from the linear region is fitted with a function following the model
described in \cite{proebst}. This  assumes an
exponential rise and two exponential decay components. 
Once the parameters of the model function are determined for the
linear region, a
mathematical description of the  changes in  pulse
shape in the non-linear range is needed. At higher energies the pulse shape  varies according to the non-linearities of the
superconducting phase transition. These changes can be described via
successive approximation to the model function.
 The higher order correction coefficients are extracted by fitting
templates of different energies up to 2.31 MeV (see \cite{methods}),
enabling  the energy
spectrum to be reconstructed up to the highest recorded $\alpha$-signal ($\alpha -\alpha$ cascade at $\sim$15 MeV).

\section{Background}

Due to the powerful
discrimination technique CRESST II detectors offer, contaminations from natural decay
chains and other $\alpha$-unstable isotopes  have been identified by
their $\alpha$-decays with a sensitivity of $\sim1 \mu$Bq/kg.
The results of this analysis have shown very good agreement with the
results obtained using different techniques commonly employed to determine
the presence of impurities in crystals.  These include
ICPMS\footnote{Gran Sasso Laboratory  and
Durham University \cite{durham}}
(Inductively Coupled Plasma Mass Spectrometry), HPGe\footnote{ Modane
Underground Laboratory and Gran
Sasso Underground Laboratory \cite{arpesella}} (High Purity
Germanium $\gamma$-spectroscopy)
and X-ray luminescence\footnote{Durham University \cite{Vitalii}} techniques.
\begin{table*}
\begin{center}
\begin{tabular}{|l|c|c|c|c|c|}
\hline
&\multicolumn{4}{c|}{Crystal B} & Crystal E\\\hline
&\multicolumn{2}{c|}{semi-}&\multicolumn{2}{c|}{}&\multicolumn{1}{c|}{semi-}\\
&\multicolumn{2}{c|}{quantitative}&\multicolumn{2}{c|}{quantitative}&\multicolumn{1}{c|}{quantitative}\\
&\multicolumn{2}{c|}{analysis}&\multicolumn{2}{c|}{analysis}&\multicolumn{1}{c|}{analysis}\\\hline
Natural&Sample I&Sample II&Sample I&Sample II&\\
Element& [ppb]& [ppb]& [ppb]&[ppb]&[ppb]\\\hline
Nd&555 $\pm$ 167&773 $\pm$ 232&683 $\pm$ 137&994 $\pm$ 199&1100 $\pm$ 330\\
Sm&$<$22&11 $\pm$ 3.3&9 $\pm$ 2&13 $\pm$ 3&$<$6\\
Eu&$<$8&4 $\pm$ 1.2&4 $\pm$ 1&5 $\pm$ 1&$<$2\\
Gd&2898 $\pm$ 869&4084 $\pm$ 1225&&&1000 $\pm$ 300\\
Er&1328 $\pm$ 398&1877 $\pm$ 566&&&940 $\pm$ 283\\
Dy&$<$15&$<$3&&&$<$3\\
Hf&$<$22&$<4$&&&$<$3\\
Os&$<$29&$<5$&&&$<$4\\
\hline
\end{tabular}
\caption{\small Results from the ICPMS analysis at the Gran Sasso facility on rare earths, hafnium,
and osmium. For the
semiquantitative analysis the calibration curve is determined with a
multi-element standard which may be different from the
one to be analysed. On some elements, in particular on samarium, a
quantitative analysis was performed. In this cases the calibration is
obtained using known standards of the element to be analysed. All the
samples are dissolved in a HNO$_3$ solution in the microwave oven with
different exposures for samples I and II.} 
\label{ICPMS}
\end{center}
\end{table*}

Evidence for contaminations due to $\alpha$-unstable rare earth
elements has been found in all the measured detectors. Rare earth doped crystals
are commonly produced for laser applications, therefore these
impurities are likely to have been introduced during the production
of the crystals. In particular an $\alpha$ peak at 2.31 MeV has been detected and
identified as the $\alpha$-decay of $^{147}$Sm. This
isotope is naturally occurring with 15$\%$ isotopic abundance and is
known to be a pure $\alpha$ unstable radionuclide with
T$_{1/2} = 1.06 \times 10^{11}$ years. The  
intensity of this peak varies for the different detectors, implying
different levels of
contamination in each crystal. Crystal B showed a count rate of (1.7
$\pm$ 0.1) counts/h which translates into
(13 $\pm$ 1) ppb of natural samarium in the crystal. The rate on the same
peak measured with  crystal E corresponds to a contamination of (5.5 $\pm$
0.4) ppb of
natural samarium. 

ICPMS analyses were performed in the Gran Sasso facility on  crystals
B and E
giving results, shown in Table \ref{ICPMS}, in agreement with our
analysis of the $\alpha$-peak intensities. Evidence of $^{144}$Nd (Q = 1905.1 keV, isotopic
abundance (i.a.) = 23.8$\%$, T$_{1/2} = 2.29 \times 10^{15}$ y) and
$^{152}$Gd (Q = 2205 keV, i.a. = 0.2$\%$, T$_{1/2} = 1.08 \times 10^{14}$
 y) \cite{isotope}
$\alpha$-decay  has also been observed and the results are in accordance with the 
ICPMS analysis.
 Furthermore, X-ray luminescence analysis on crystal B showed the presence of
the characteristic emission peak of Gd and Er \cite{ninkovic}, the
two most abundant rare earth contaminants (see Table \ref{ICPMS}).

Finally, alternative candidates such as $^{174}$Hf (Q =2496 keV, i.a. =
0.162$\%$, T$_{1/2} = 2.0 \times 10^{15}$ y)  and $^{186}$Os (Q = 2822
keV, i.a. = 1.58$\%$, T$_{1/2} = 2.0 \times 10^{15}$ y)\cite{isotope}
were considered to cause the 2.31 MeV peak. They were
discarded, however, due to the unreasonably high atomic concentration needed
to produce the detected rate.  

\section{Calibration}
\begin{figure*}[!ht] 
\includegraphics[height=9cm]{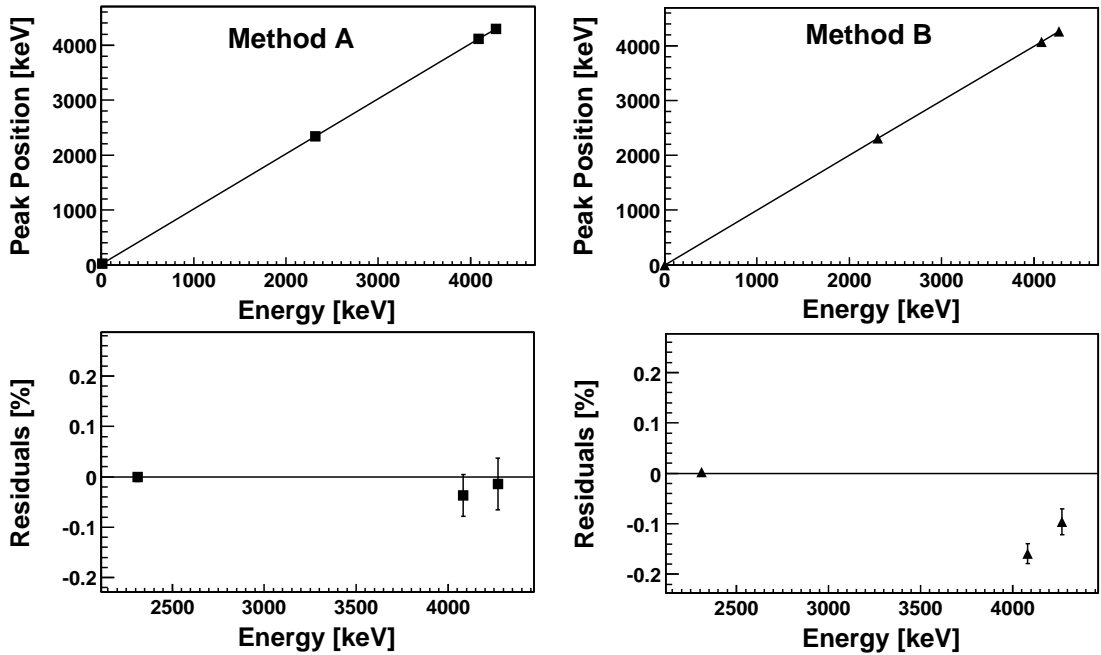}
\caption{\small Top: Fitted peak position versus known energy. The
energy scale for the y-axis is  fixed by the $^{147}$Sm peak while the position of
the $^{232}$Th and $^{238}$U peaks results from a simple linear
extrapolation of the calibration for both methods A and B. 
Bottom:  Residuals of
the fitted position from the expected position. For method A, the
$^{232}$Th peak  results at (4081.3 $\pm$ 1.7) keV ($\Delta$E = (28
$\pm$ 4) keV (FWHM)) and the
$^{238}$U at  (4269.4 $\pm$ 2.2) keV. For method B, the peaks lie
 at (4076.3 $\pm$ 0.8) keV ($\Delta$E = (14.1 $\pm$ 1.7) keV (FWHM)) and
(4265.9 $\pm$ 1.1) keV respectively. The calibration of the amplitude
spectrum resulting from method A is much more accurate; however, the resolution is less compared with method B.
}
\label{accuracy}
\end{figure*}

Once unambiguously identified, $^{147}$Sm was used to calibrate the
full $\alpha$ spectrum. This allowed a precise identification of the U-Th
peaks, which were then  used to determine the accuracy of the
calibration for both reconstruction methods A and B.
In Fig. \ref{accuracy} the linearity of the response functions
for method A (top left figure) and method B (top right figure) is
shown for the data of run 28. On the x-axis the central values of the tabulated energies
\cite{wapstra} of the $^{147}$Sm (Q= 2310.5 $\pm$ 1.1 keV),
$^{232}$Th (Q= 4082.8 $\pm$ 1.4 keV) and
$^{238}$U (Q= 4270 $\pm$ 3 keV) peaks are given. On the y-axis their fitted
peak position is plotted.  The energy
scale for the y-axis is fixed on the $^{147}$Sm peak at 2310.5 keV while the position of
the $^{232}$Th and $^{238}$U peaks results from a simple linear
extrapolation of the calibration. 
At the
bottom of the figure the residuals of the fitted position with respect to the expected position are shown. 
The calibration of the amplitude spectrum resulting from method A is
much more accurate, but
less information can be used to reconstruct the signal amplitude compared with method B.
This results in a slightly worse energy resolution $\Delta$E =
(12.9 $\pm$ 0.3) keV (FWHM) on the
$^{147}$Sm peak, and it fails to
completely reconstruct the higher part of the U-Th $\alpha$-spectrum.
Method B, on the other hand, using a larger amount of  information from the pulses than
method A,  gives a much better resolution ($\Delta$E = (6.7 $\pm$ 0.1)
keV (FWHM) on the $^{147}$Sm peak), but the approximation of the pulse shape description introduces a
higher systematic error on the linear extrapolation of the calibration. 
In the following, method B has been used to fully reconstruct
and identify the entire $\alpha$ spectrum, while method A was applied as
an independent check on the peak position of the $^{180}$W evidence.
In particular, method A is used to extract the measured energy for  the
$^{180}$W Q-value.

\section{Results}
\begin{figure*}
\begin{center}
\includegraphics[height=9cm]{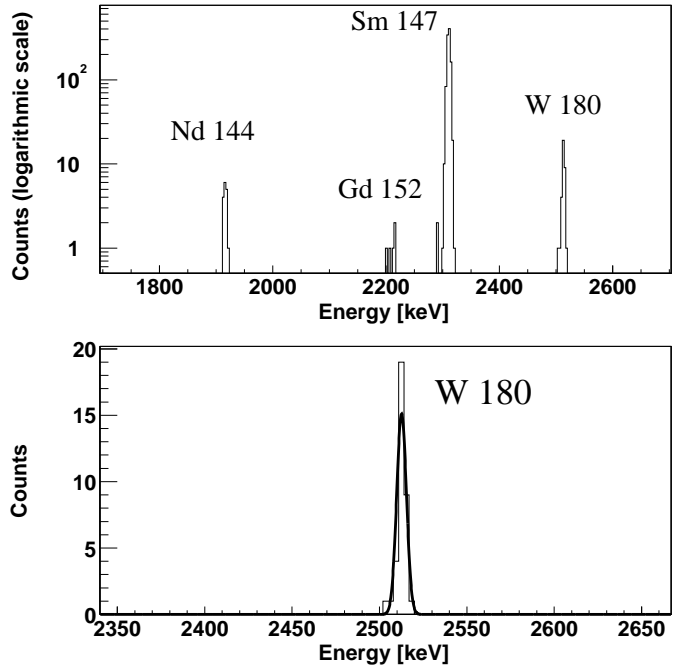}
\caption{\small Alpha events measured in Run 28. In the top figure the
energy spectrum between 1.7 and 2.7 MeV is shown with 3 keV
bins. On the  vertical axis the number of counts are displayed on a
logarithmic scale.
In the bottom figure the fit to the $^{180}$W $\alpha$-decay
is shown. No background counts are present in a wide energy interval around the peak. }
\label{result}
\end{center}
\end{figure*}

Crystal D showed the best performance as a
detector (energy resolution: $\Delta$E = 1.8 keV (FWHM) for
122 keV gammas, and $\Delta$E =(6.7 $\pm$ 0.1) keV for 2.31 MeV alphas, using
method B) and it has the longest
exposure (11.745 kg days in run 27 and 12.268 kg days in run 28). In run 28
the performance of the module was improved by replacing the Ag 
reflective foil used in run 27, which showed a $^{210}$Pb contamination,
by a highly reflective (99$\%$ at 420 nm) scintillating polymeric foil. It
is these data  which we discuss in full detail. 
A peak was
observed at  2512.8 $\pm$ 0.4 (stat.) $\pm$ 4.3 (sys.) keV with
a resolution  $\Delta$E = (6.5 $\pm$ 0.7) keV (FWHM). This is shown in
Fig. \ref{result}.
The position in energy obtained by reconstructing the spectrum with method A
is (2516.5 $\pm$ 1.4) keV.
A half-life
T$_{1/2}=(1.7 \pm 0.2) \times 10^{18}$ y is obtained. This is consistent
with the value from run 27 on the same crystal  and
on crystals B and E, as shown in Table \ref{halflife}.
The consistency of the results from different crystals tends to confirm  that
the signal originates from the $^{180}$W decay and not from unknown impurities which may
be expected to vary for each individual crystal, as seen in Section 4. Furthermore ICPMS
measurements performed on these two crystals reported in Table \ref{ICPMS}
exclude alternative explanations for the tungsten peak, such as
$^{174}$Hf.
\begin{table*}
\begin{center}
\begin{tabular}{|lllll|}
\hline
Crystal&Run &Exposure&Number&Half-life\\
&number &[kg days]&of counts&[y]\\\hline
D& 28& 12.268 &35$ \pm$ 4.3&T$_{1/2} = (1.7 \pm 0.2) \times 10^{18}$\\
D& 27& 11.745&28.5 $\pm$ 5.6&T$_{1/2} = (1.9\pm0.4) \times 10^{18}$\\
E& 23& 3.467&9&T$_{1/2} \geq 1.1 \times 10^{18}$ (90$\%$
C.L.)\\
B& 22& 1.14 &4&T$_{1/2} \geq 6.8 \times 10^{17}$ (90$\%$
C.L.)\\\hline
\hline
\multicolumn{2}{|c}{All Crystals}& 28.62 &75.5$\pm$8.7&T$_{1/2} = (1.8
\pm 0.2) \times 10^{18}$\\\hline
 \end{tabular}
 \caption{Half-life for the $\alpha$-decay of $^{180}$W obtained
from the present work. The results are given independently for the
four different runs and for the total exposure. For the runs 27, 28,
and the total exposure, the number of counts is determined by a
Gaussian fit to the peak. Errors are given at 1$\sigma$. For run 22 and 23, due to the small number of
counts, the total number of events contained within $\pm 3\sigma$ of the
expected energy position is given. }
\label{halflife} 
 \end{center}
\end{table*}

\begin{figure}
\begin{center}
\includegraphics[height=4cm]{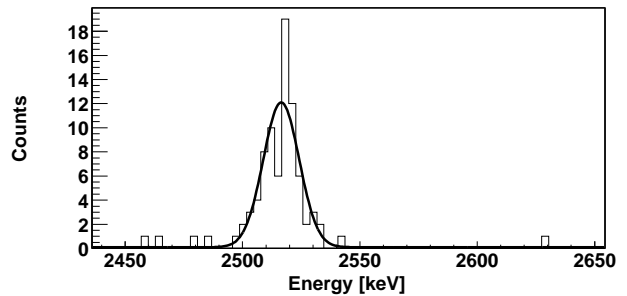}
\caption{\small Sum of the four different runs, each analysed
independently with method A. The resulting peak is fitted giving an energy
of (2516.4 $\pm$ 1.1) keV with
resolution FWHM = (18 $\pm$ 2) keV. Few background events due to a surface
contamination of the Ag foil  used in
run 27 are visible. The scintillating foil
used in runs 22, 23 and 28  efficiently vetoes all these events (see text).}
\label{sum}
\end{center}
\end{figure}

Finally, the four measurements were added, resulting in a spectrum for
a total exposure of
28.62 kg days. The data from each run have been analysed with method
A, calibrated individually and then summed. This results in half-life of
T$_{1/2} = (1.8 \pm 0.2) \times 10^{18}$y, which is consistent  with the
previously published limits. Including the uncertainty on the position of the $^{147}$Sm line, a peak
of energy (2516.4 $\pm$ 1.1 (stat.) $\pm$ 1.2 (sys.)) keV is found with a
resolution $\Delta$E = (18 $\pm$ 2) keV (FWHM), as shown in
Fig. \ref{sum}.
 This result is consistent with expectations based on the
mass difference of $^{180}$W  and $^{176}$Hf from
Ref.\cite{wapstra}, but lies  2$\sigma$ away from the latest
update \cite{wapstra03} (see Table \ref{Qvalues}). We note however that
 most of the input data used 
to evaluate the atomic masses consist of relative measurements which set
a relation in mass or energy among two or more nuclei. In the diagram
showing the relations among input data (Fig 1 of Ref.\cite{wapstra03}), $^{180}$W and $^{176}$Hf are
only weakly connected and an underestimation of the errors is
plausible. Our result represents the missing direct connection between $^{180}$W and $^{176}$Hf. 

Limits on the half-lives for the non-observed $\alpha$-decays of the
other four naturally occurring tungsten isotopes were also calculated. For this purpose the data from
run 27 could not be used. In that run, a Ag foil was used to surround
the crystal, while in the other three runs the scintillating polymeric
foil was used. Any 
contamination of $\alpha$-unstable isotopes, such as $^{210}$Po on the
surface of the foil or on the crystal may results in $\alpha$ events of degraded energy from
the U-Th chain energy,  going down to $\sim$100 keV. These were observed
in run 27 with a rate of 0.087
counts/100keV/kg/day both between $\sim$500 keV and the $^{144}$ Nd
peak at 1.9 MeV, and between the $^{180}$W and the $^{232}$Th peaks. The polymeric foil
used in all the other runs did not show any surface
contamination. Furthermore if the scintillating side of the foil faces
directly the crystal, as in run 28, it efficiently vetoes all the surface events  due to
the light emitted by the polymeric foil itself. 
The limits obtained from the sum of runs 22, 23 and 28 are reported in
Table \ref{lim}.

\begin{table}
\begin{center}
 \begin{tabular}{|llll|}
\hline
Isotope&\multicolumn{3}{c|}{Half-life [y]}\\
&This work&Previous \protect{\cite{kiev}}&Previous
\protect{\cite{cebrian}} \\\hline 
$^{182}$W&T$_{1/2} \geq 7.7 \times10^{21}$&$\geq
1.7 \times 10^{20}$&$\geq
2.5 \times 10^{19}$\\
$^{183}$W&T$_{1/2} \geq 4.1 \times 10^{21}$&$\geq
0.8 \times 10^{20}$& $\geq
1.3 \times 10^{19}$\\
$^{184}$W&T$_{1/2} \geq 8.9 \times 10^{21}$& $\geq
1.8 \times 10^{20}$& $\geq
2.9 \times 10^{19}$\\
$^{186}$W&T$_{1/2} \geq 8.2 \times 10^{21}$&$ \geq
1.7 \times 10^{20}$&$\geq
2.7 \times 10^{19}$\\ 
\hline
\end{tabular}
\caption{Lower limits (90$\%$ C.L.) on the half-lives of  the non-observed $\alpha$-decays of the
other four naturally occurring tungsten isotopes, obtained with a total exposure of 16.875
kg days. For comparison the best previously
published limits are also shown \protect{\cite{kiev, cebrian}}.}
\label{lim} 
\end{center}
\end{table}

\section{Conclusions}
 We have observed the natural $\alpha$-decay of $^{180}$W with  a
half-life of T$_{1/2} = (1.8 \pm 0.2) \times 10^{18}$y and a Q-value of
 (2516.4 $\pm$ 1.1 (stat.) $\pm$ 1.2 (sys.)) keV.   
In addition, the lack of any signal
from the other tungsten isotopes sets new limits
on their half-lives. These results offer an improvement of roughly
a factor of 50 over the best previously published limits.  
Finally we would like to note the power and potential of this cryogenic
technique, which here enabled us to achieve a very high sensitivity in an application to nuclear physics.

\section*{Acknowledgements}
We would like to thank S. Nisi and M. Balata for the ICPMS measurements and
M. Laubenstein for the $\gamma$ spectroscopy analysis  at the Gran
Sasso Laboratories. Thanks also to C. Goldbach
for the Ge measurements at Modane and to the University of Durham for
the ICPMS and X-ray luminescence analysis. 
This work was supported by PPARC, BMBF, the EU Network
HPRN-CT-2002-00322 on Applied Cryodetectors, the EU Network on
Cryogenic Detectors (contract ERBFMRXCT980167), the DFG SFB 375 on
Particle Astrophysics and two EU Marie Curie Fellowships.

\end{document}